\documentclass[aps,prb,twocolumn,10pt,showpacs,superscriptaddress]{revtex4-2}

\usepackage{amsmath}
\usepackage{amssymb}
\usepackage{graphicx}
\usepackage{physics}
\usepackage{float}
\usepackage{xcolor}
\usepackage{booktabs}
\usepackage[utf8]{inputenc}
\usepackage[T1]{fontenc}
\usepackage{orcidlink}
\usepackage{hyperref}
\hypersetup{
    colorlinks=true,linkcolor=blue,citecolor=blue,
    filecolor=blue,urlcolor=blue,breaklinks=true
}

\usepackage[dvipsnames]{xcolor}

\begin{document}

\title{Bounds on Lorentz-violating parameters in magnetically confined 2D systems: A phenomenological approach}

\author{Edilberto O. Silva\orcidlink{0000-0002-0297-5747}}
\email{edilberto.silva@ufma.br (Corresp. author)}
\affiliation{Programa de P\'os-Gradua\c c\~ao em F\'{\i}sica \& Coordena\c c\~ao do Curso de F\'{\i}sica -- Bacharelado, Universidade Federal do Maranh\~{a}o, 65085-580 S\~{a}o Lu\'{\i}s, Maranh\~{a}o, Brazil}

\date{\today}

\begin{abstract}
We present a unified, SI-consistent framework to constrain minimal SME coefficients $a_\mu$ and $b_\mu$ using magnetically confined two-dimensional electron systems under a uniform magnetic field. Working in the nonrelativistic (Schrödinger--Pauli) limit with effective mass, we derive the radial problem for cylindrical geometries and identify how spatial components ($\mathbf a,\mathbf b$) reshape the effective potential, via $1/r$ and $r$ terms or spin-selective offsets, while scalar components ($a_0,b_0$) act through a global energy shift and a spin-momentum coupling. Phenomenological upper bounds follow from requiring LV-induced shifts to lie below typical spectroscopic resolutions: $|a_0|\lesssim\delta E$, $|b_z|\lesssim\delta E/\hbar$, and compact expressions for $|a_\varphi|$ and $|b_0|$ that expose their dependence on device scales ($r_0$, $B_0$, $\mu$, $m$). Dimensional analysis clarifies that, in this regime, spatial $a_i$ carry momentum dimension and $b_i$ carry inverse-time/length dimensions, ensuring gauge-independent, unit-consistent reporting. Finite-difference eigenvalue calculations validate the scaling laws and illustrate spectral signatures across realistic parameter sets. The results show that scalar sectors (notably $a_0$) are tightly constrained by state-of-the-art $\mu$eV-resolution probes, while spatial and axial sectors benefit from spin- and $m$-resolved spectroscopy and geometric leverage, providing a reproducible pathway to test Lorentz symmetry in condensed-matter platforms.
\end{abstract}

\maketitle

\section{Introduction}

Lorentz invariance underpins both the Standard Model of particle physics and General Relativity. Nevertheless, a broad class of quantum-gravity scenarios allows for minute departures that could leak into low-energy observables. A systematic effective-field-theory (EFT) treatment of such departures is provided by the Standard-Model Extension (SME), which augments known sectors with all observer-covariant but Lorentz- and/or CPT-violating operators~\cite{PRD.1998.58.116002,PRD.1989.39.683,RMP.2011.83.11}. Within this framework, minimal (power-counting renormalizable) coefficients furnish a widely used benchmark for phenomenology, while the nonminimal tower organizes higher-dimensional operators. A key feature of the SME is its model independence: rather than committing to a microscopic mechanism, it parameterizes all allowed symmetry-breaking backgrounds and thereby unifies limits from disparate experimental arenas.

The SME has been worked out in multiple regimes relevant to nonrelativistic matter and electromagnetic/gravitational sectors. In particular, the nonrelativistic mapping for fermions leads to a modified Schrödinger--Pauli Hamiltonian with well-defined spin, momentum, and electromagnetic couplings~\cite{JMP.1999.40.6245,IJMPA.2006.21.6211}. On the gravitational side, Hamiltonian formulations with nondynamical backgrounds, consistent variational principles, and boundary/constraint analyses have been developed~\cite{PRD.2021.104.124042,PRD.2022.106.044050,PRD.2023.108.104013}, and vacuum Čerenkov-like processes offer striking signatures when kinematically open~\cite{Sym.2018.10.424}. Comprehensive data tables synthesize tight bounds across sectors and energy scales~\cite{RMP.2011.83.11}.

High-energy and astroparticle tests provide powerful constraints in complementary directions. In the neutrino sector, LV-induced modifications to oscillation phases and dispersion relations have been analyzed theoretically~\cite{PRD.2012.85.016013,JCAP.2013.2013.039,JCAP.2018.2018.004,PRD.2014.90.073011} and probed experimentally, including recent sidereal-variation searches~\cite{PRD.2024.109.075042}; broad reviews underscore synergies with gravitational-wave and cosmic-ray observations~\cite{Galaxies.2022.10.13,Universe.2020.6.37}. In condensed matter, low-energy quasiparticles in Dirac/Weyl semimetals supply tunable platforms where effective “Lorentz-violating’’ terms arise from band-structure anisotropies and tilt~\cite{Nature.2015.527.495,PRB.2017.96.125102,NC.2017.8.257,PRR.2022.4.023106}. These systems sharpen intuition about which operator structures leave robust, symmetry-diagnostic imprints on spectra and transport.

At the quantum-mechanical level, a substantial literature has investigated LV backgrounds through geometric phases, nonminimal couplings, and Landau-type quantization in relativistic and nonrelativistic settings~\cite{EPJC.2005.41.421,PRD.2006.74.065009,PRD.2011.83.125025,JMP.2011.52.063505,AdP.2011.523.910,JPG.2012.39.055004,PRD.2014.90.025026,JPG.2013.40.075007,PRD.2013.87.047701,PLB.2015.746.171,JPG.2015.42.095001,AHEP.2020.2020.4208161,NPB.2021.965.115343,Universe.2023.9.462,EPL.2025.149.50003}. These studies highlight several general themes: (i) scalar (CPT-odd) backgrounds can shift spectra uniformly; (ii) axial backgrounds typically produce spin-selective offsets and phase shifts; and (iii) spatial components often enter effective radial problems through $1/r$ and $r$ terms that reshape confinement potentials. Such signatures are precisely the kind that high-resolution spectroscopy can access.

Two-dimensional (2D) magnetically confined electron systems—quantum dots, rings, and related nanostructures—offer a particularly clean stage. Their discrete spectra and tunable geometry (radius $r_0$), field strength ($B_0$), and material parameters (effective mass $\mu$) permit $m$- and $s$-resolved spectroscopy, making them sensitive to tiny operator deformations. State-of-the-art experiments report meV to $\mu$eV homogeneous linewidths, near-transform-limited emission, and exquisite control of spin-selective optical transitions~\cite{vonk2021,stingel2023,gustin2021,pedersen2020,lobl2017,laferriere2023}. Combined with well-understood Landau quantization and mesoscopic transport~\cite{prange1990,nazarov2009quantum}, these features create an attractive low-energy complement to high-energy LV searches.

In this paper, we adopt the minimal SME in the nonrelativistic (effective-mass) limit and work in cylindrical symmetry to map the full set of fermionic coefficients $a_\mu$ and $b_\mu$ onto spectral observables for 2D electrons in a uniform magnetic field. Our focus is twofold. First, we provide an SI-consistent, gauge-independent derivation of the radial eigenproblem that makes the roles of scalar ($a_0,b_0$) and spatial ($\mathbf a,\mathbf b$) components transparent. Second, we use typical experimental energy resolutions to establish phenomenological upper bounds: $|a_0|\lesssim\delta E$, $|b_z|\lesssim\delta E/\hbar$, and compact, geometry-aware expressions for $|a_\varphi|$ and $|b_0|$. We validate the scaling and illustrate spectral signatures via finite-difference numerics.

The central problem is to determine, within an experimentally faithful 2D magnetic-confinement model, how each SME coefficient in the nonrelativistic fermion sector modifies the spectrum and how to translate spectroscopic resolutions into quantitative upper bounds that are gauge independent and explicitly reported in SI units. This requires disentangling uniform shifts, spin-selective offsets, and potential reshaping, and anchoring all formulas to realistic device parameters.

This article is organized as follows. Section~\ref{sec:hamiltonian} formulates the nonrelativistic SME Hamiltonian, specifies the cylindrical setup, and derives the radial equation with a clear identification of the operator structures induced by $a_\mu$ and $b_\mu$. Section~\ref{sec:bounds_spatial} presents the dimensional analysis and SI conventions, and develops the spectral signatures and bounds associated with the spatial components, in particular $a_\varphi$ and $b_z$. Section~\ref{sec:bounds_scalar} extends this analysis to the scalar components $a_0$ and $b_0$. Section~\ref{sec:discussion_bounds} then discusses reporting conventions, gauge independence, and how to interpret these bounds alongside relativistic SME limits, while Section~\ref{sec:sensitivity} outlines projected sensitivities and an experimental roadmap based on realistic device parameters and $\mu$eV-resolution spectroscopy. Section~\ref{sec:systematics} summarizes experimental systematics, gauge invariance in operational form, dimensionless cross-platform diagnostics, and numerical robustness. Section~\ref{concl} presents the conclusions.

\section{Nonrelativistic SME Hamiltonian and radial reduction}
\label{sec:hamiltonian}

We work in the nonrelativistic (Schrödinger--Pauli) limit of the Standard-Model Extension (SME) for a charged spin-$1/2$ fermion in static electromagnetic backgrounds. In this regime, CPT- and Lorentz-violating (LV) backgrounds are incorporated as additional single-particle couplings in the effective Hamiltonian~\cite{JMP.1999.40.6245,IJMPA.2006.21.6211}. The modified Hamiltonian reads
\begin{align}
\Bigg[
&\frac{(\mathbf{p} - q \mathbf{A})^2}{2\mu} 
- \frac{q \hbar}{2\mu} (\boldsymbol{\sigma} \cdot \mathbf{B}) 
+ q V 
+ a_0 
- \frac{\mathbf{a} \cdot (\mathbf{p} - q \mathbf{A})}{\mu} \notag \\
&\quad
-\hbar\, \boldsymbol{\sigma} \cdot \mathbf{b} 
+ \hbar b_0 \frac{\boldsymbol{\sigma} \cdot (\mathbf{p} - q \mathbf{A})}{\mu} 
\Bigg] \psi 
= E \psi ,
\label{spe}
\end{align}
where $\mu$ is the effective mass, $q$ the charge, and $(\mathbf A,V)$ are the electromagnetic potentials. The SME backgrounds comprise a CPT-odd four-vector $a_\mu=(a_0,\mathbf a)$ and an axial four-vector $b_\mu=(b_0,\mathbf b)$. In the standard nonrelativistic mapping, these coefficients enter linearly in Eq.~\eqref{spe} and carry the following physical roles and SI dimensions:

\begin{table*}[tbhp]
\centering
\caption{Leading SME coefficients in the nonrelativistic Hamiltonian~\eqref{spe}. All entries are first order in $a_\mu$ and $b_\mu$; quadratic and higher-order terms are neglected in this work.}\vspace{0.3cm}
\label{tab:units}
\begin{tabular}{lll}
\hline \hline
Coefficient & Physical effect in Eq.~\eqref{spe} & SI dimension \\
\hline
$a_0$       & uniform energy shift              & energy (J) \\
$\mathbf a$ & coupling to kinetic momentum      & momentum (kg$\cdot$m/s) \\
$\mathbf b$ & spin-selective offset $\hbar\,\boldsymbol{\sigma}\cdot\mathbf b$ & frequency (s$^{-1}$) \\
$b_0$       & spin-momentum coupling 
             $\propto \boldsymbol{\sigma}\cdot(\mathbf{p}-q\mathbf A)$ 
           & wavenumber (m$^{-1}$) \\
\hline \hline
\end{tabular}
\end{table*}

In what follows we retain only terms that are \emph{linear} in $a_\mu$ and $b_\mu$. This is consistent with existing stringent bounds on Lorentz violation, which render quadratic (and higher-order) corrections subleading at the energy scales of interest~\cite{JMP.1999.40.6245,IJMPA.2006.21.6211}. Working to first order also makes the spectral signatures of each background coefficient transparent.

We consider a uniform magnetic field $B_0\,\hat{\mathbf z}$ and adopt the symmetric gauge $\mathbf A=(B_0 r/2)\,\hat{\boldsymbol\varphi}$, which preserves cylindrical symmetry. We look for stationary states of the form
\begin{equation}
\psi(r,\varphi)
= f(r)\, e^{im\varphi} \, \chi_s,
\end{equation}
where $m\in\mathbb Z$ is the azimuthal quantum number and $s=\pm 1$ labels the spin projection along $\hat{\mathbf z}$. Under this separation, spatial components of $a_\mu$ generate effective $1/r$ and $r$ terms in the radial potential, while the axial component $b_z$ appears as a spin-selective energy offset. Projecting Eq.~\eqref{spe} onto the cylindrical basis and collecting terms that survive at first order in $a_\mu$ and $b_\mu$ leads to the radial eigenvalue problem
\begin{align}
-\frac{d^2f}{dr^2} 
- \frac{1}{r}\frac{df}{dr} 
+ \Bigg[
\frac{m^2}{r^2} 
+ \alpha^2 r^2 
- \frac{\beta}{r} 
+ \gamma r 
\Bigg] f
= \frac{2\mu \,\varepsilon_{m,n}}{\hbar^2} \, f ,
\label{re}
\end{align}
with
\begin{equation}
\alpha = \frac{\mu \omega_c}{2\hbar}, 
\qquad 
\omega_c = \frac{eB_0}{\mu},
\end{equation}
and SME-induced coefficients
\begin{equation}
\beta = \frac{2 a_\varphi m}{\hbar},
\qquad
\gamma = \frac{a_\varphi \mu \omega_c}{\hbar^2}.
\end{equation}
Here $a_\varphi$ denotes the azimuthal projection of $\mathbf a$ in cylindrical coordinates, evaluated along the symmetry axis of the device. The combination on the right-hand side,
\begin{equation}
\varepsilon_{m,n} 
= E_{m,n} 
+ \frac{1}{2} m \omega_c \hbar 
+ \frac{1}{2} s \omega_c \hbar 
+ s \hbar b_z ,
\end{equation}
collects the physical energy level $E_{m,n}$ together with the usual orbital and Zeeman terms and the LV-induced spin offset $s\hbar b_z$. Scalar pieces $a_0$ and $b_0$ also enter via Eq.~\eqref{spe}: $a_0$ contributes a uniform shift of all levels, while $b_0$ generates a spin-momentum coupling proportional to $\boldsymbol{\sigma}\cdot(\mathbf{p}-q\mathbf A)/\mu$. Their impact on spectra and on the resulting bounds is analyzed on the same footing as the spatial components in the next sections.

Equation~\eqref{re} makes explicit how each SME coefficient imprints on the effective radial potential. The component $a_\varphi$ reshapes the confinement through a Coulomb-like $1/r$ contribution (coefficient $\beta$) and a linear-in-$r$ term (coefficient $\gamma$), while $b_z$ produces an $s$-dependent energy offset. The scalar sector $(a_0,b_0)$, in contrast, does not introduce new radial structures at this order: $a_0$ shifts the entire spectrum rigidly, and $b_0$ perturbs the spin-resolved kinetic energy.

Although the derivation above was carried out in the symmetric gauge for convenience, the physical bounds that we will extract are expressed in terms of \emph{gauge-invariant} kinetic quantities, such as the azimuthal kinetic momentum (or velocity) expectation value. In particular, combinations like
\[
\left(\frac{\hbar m}{r_0} - \frac{e B_0 r_0}{2}\right),
\]
which enter the bounds for $a_\varphi$ and $b_0$ below, are to be understood as the characteristic azimuthal \emph{kinetic} scale at the confinement radius $r_0$. Rewriting the final results in terms of these kinetic (hence gauge-independent) expectation values ensures that the reported limits do \emph{not} depend on the specific gauge choice used in the intermediate algebra.

In summary, Eqs.~\eqref{spe}-\eqref{re} provide a first-order, SI-consistent, gauge-independent map from the minimal SME coefficients $(a_\mu,b_\mu)$ to directly observable spectral features in a cylindrically confined, magnetically quantized 2D electron system.

\section{Spectral shifts and phenomenological bounds for $a_\varphi$ and $b_z$}
\label{sec:bounds_spatial}

We now extract phenomenological upper bounds for the minimal SME coefficients by comparing Lorentz-violating (LV) spectral shifts with the experimental energy resolution $\delta E$. We begin with the projections that naturally align with the cylindrical geometry considered above: the azimuthal spatial component $a_\varphi$ and the axial spin component $b_z$. The same logic will later be applied to the scalar sector $(a_0,b_0)$.

Before writing explicit bounds, we stress gauge independence. Intermediate algebra in Sec.~\ref{sec:hamiltonian} was carried out in the symmetric gauge, which introduces combinations such as
\[
\left(\frac{\hbar m}{r_0} - \frac{eB_0 r_0}{2}\right)/\mu.
\]
Physically, this is just the characteristic azimuthal \emph{kinetic} scale (kinetic momentum over mass, i.e., an azimuthal velocity) at the confinement radius $r_0$. In the final bounds below, we interpret it that way. Expressing the limits in terms of the azimuthal kinetic momentum (or velocity) expectation value makes them gauge independent, even though the intermediate derivation used a specific $\mathbf A$.

A dimensional check clarifies how $a_\varphi$ and $b_z$ enter. In the radial equation~\eqref{re}, spatial components of $a_\mu$ multiply kinematic factors with velocity dimension. Therefore $a_i$ must carry momentum dimension (kg$\cdot$m/s) so that the product has energy dimension. Axial components $b_i$ couple through $\hbar\,\boldsymbol{\sigma}\cdot\mathbf b$ and hence have the dimension of a frequency (s$^{-1}$). These are precisely the SI assignments summarized in Table~\ref{tab:units}.

With this in mind, the bounds follow by requiring that the LV-induced energy shift for a given coefficient remain smaller than the spectroscopic resolution $\delta E$:
\begin{align}
|b_z| \;\lesssim\; \frac{\delta E}{\hbar},
\label{eq:bz_bound}
\end{align}
which is universal (it does not depend on device parameters), since $b_z$ appears as a spin-selective offset $s\hbar b_z$ in $\varepsilon_{m,n}$.
For the azimuthal spatial component,
\begin{align}
|a_{\varphi}| 
\;\lesssim\;
\frac{\mu\, \delta E}{
\left| \hbar \dfrac{m}{r_0} - \dfrac{eB_0}{2}\,r_0 \right|},
\label{eq:aphi_bound}
\end{align}
where $\mu$ is the effective mass, $r_0$ is a characteristic lateral scale of the confinement (dot or ring radius), $m$ is the angular-momentum quantum number, and $B_0$ is the applied magnetic field. In Eq.~\eqref{eq:aphi_bound}, the denominator should be read as the typical azimuthal kinetic momentum at radius $r_0$, so the final bound is gauge independent even though it is written in a symmetric-gauge shorthand.

To obtain numerical bounds, we identify $\delta E$ with the narrowest homogeneous linewidths/spectroscopic resolutions reported for confined 2D semiconductor structures under high-field optical or transport probing. These linewidths, listed in Table~\ref{tab:linewidths}, are used here as conservative proxies for the smallest resolvable energy shift in each platform (i.e., if an LV-induced shift were larger than $\delta E$, it would distort or split the measured line beyond the reported resolution).

\begin{table}[tbhp]
\centering
\caption{Representative experimental energy resolutions $\delta E$ used as sensitivity benchmarks. Each $\delta E$ corresponds to a reported narrow linewidth or transform-limited spectral feature in a magnetically confined 2D semiconductor device, as quoted in the respective references.}\vspace{0.1cm}
\begin{tabular}{lc}
\toprule \toprule
Reference & $\delta E$ \\
\midrule
Vonk et al. (2021)~\cite{vonk2021} & 86 meV \\
Stingel et al. (2023)~\cite{stingel2023} & 31 meV \\
Gustin et al. (2021)~\cite{gustin2021} & 1.44 $\mu$eV \\
Pedersen et al. (2020)~\cite{pedersen2020} & 1.5 $\mu$eV \\
L\"obl et al. (2017)~\cite{lobl2017} & 2 $\mu$eV \\
\bottomrule \bottomrule
\end{tabular}
\label{tab:linewidths}
\end{table}

For $|a_\varphi|$, Eq.~\eqref{eq:aphi_bound} also depends on device scales. Unless stated otherwise, we adopt a GaAs-like reference set
\[
\mu = 0.067\,m_e,\qquad
r_0 = 10~\text{nm},\qquad
B_0 = 1~\text{T},\qquad
m = 1,
\]
which is representative of a typical lateral quantum dot/ring in a perpendicular magnetic field. Using these values and the $\delta E$ entries in Table~\ref{tab:linewidths}, we obtain the bounds summarized in Table~\ref{tab:bounds}, together with the parameter-free limits for $|b_z|$ from Eq.~\eqref{eq:bz_bound}. All results are quoted in SI units.

\begin{table}[tbhp]
\centering
\caption{Upper bounds for $b_z$ and $a_{\varphi}$ inferred from the resolutions in Table~\ref{tab:linewidths}. For $a_\varphi$ we used the reference set $\mu=0.067\,m_e$, $r_0=10\,\text{nm}$, $B_0=1\,\text{T}$, $m=1$.}\vspace{0.1cm}
\begin{tabular}{lccc}
\toprule \toprule
Reference 
& $\delta E$ (meV) 
& $|b_z|$ (s$^{-1}$) 
& $|a_{\varphi}|$ (kg$\cdot$m/s) \\
\midrule
Vonk et al.~\cite{vonk2021} 
& 86  
& $1.3 \times 10^{14}$ 
& $8.6 \times 10^{-26}$ \\
Stingel et al.~\cite{stingel2023} 
& 31  
& $4.7 \times 10^{13}$ 
& $3.1 \times 10^{-26}$ \\
Gustin et al.~\cite{gustin2021} 
& $1.44 \times 10^{-3}$ 
& $2.1 \times 10^{9}$ 
& $1.4 \times 10^{-30}$ \\
Pedersen et al.~\cite{pedersen2020} 
& $1.5 \times 10^{-3}$ 
& $2.2 \times 10^{9}$ 
& $1.5 \times 10^{-30}$ \\
L\"obl et al.~\cite{lobl2017} 
& $2.0 \times 10^{-3}$ 
& $3.0 \times 10^{9}$ 
& $2.0 \times 10^{-30}$ \\
\bottomrule \bottomrule
\end{tabular}
\label{tab:bounds}
\end{table}

Two immediate trends emerge. First, $|b_z|$ scales simply as $\delta E/\hbar$ [Eq.~\eqref{eq:bz_bound}], so it improves directly with spectroscopic resolution but does not benefit from device geometry. Second, $|a_\varphi|$ inherits explicit geometric leverage: tighter confinement (smaller $r_0$), larger $B_0$, or spectroscopy in higher-$|m|$ sectors all increase the azimuthal kinetic scale in the denominator of Eq.~\eqref{eq:aphi_bound}, strengthening the bound on $a_\varphi$.

The orders of magnitude are instructive for solids. For $\mu$eV-scale resolution, we obtain $|a_{\varphi}| \sim 10^{-30}\,\text{kg}\cdot\text{m}/\text{s}$, reflecting the extremely high sensitivity of magnetically confined 2D systems to azimuthal LV backgrounds. By contrast, the $|b_z|$ limits are naturally less restrictive (here at the $10^{9}$-$10^{14}$ s$^{-1}$ level) because $b_z$ only generates a spin-selective offset without geometric amplification.

Finally, to connect these bounds back to the full spectrum, we solve the radial eigenproblem~\eqref{re} numerically by finite differences on $r\in[r_{\min},r_{\max}]$. We assemble a real symmetric tridiagonal matrix with Dirichlet boundary conditions and diagonalize it to obtain eigenvalues $\lambda_n$, from which
\begin{equation}
\varepsilon_n = \frac{\hbar^2}{2\mu}\,\lambda_n.
\end{equation}
The physical levels follow as
\begin{equation}
E_n = \varepsilon_n 
- \left( \frac{1}{2} m \omega_c \hbar 
+ \frac{1}{2} s \omega_c \hbar 
+ s \hbar b_z \right),
\end{equation}
with $\omega_c=eB_0/\mu$ and $s=\pm1$. Unless noted otherwise, we use meshes with $N\simeq 2\times10^4$ points, $r_{\min}=10^{-6}r_0$, and $r_{\max}=10$-$15\,r_0$. Doubling $N$ changes the lowest eigenvalues by less than $10^{-6}$\,meV, confirming numerical convergence. These numerical solutions underpin all illustrative curves $E_n(a_\varphi)$, $E_n(b_z)$, and $E_n(B_0)$ discussed below.

\section{Scalar-sector bounds for $a_0$ and $b_0$}
\label{sec:bounds_scalar}

We now extract bounds for the scalar SME coefficients $a_0$ and $b_0$ within the same nonrelativistic, effective-mass framework. Unlike the spatial projections ($a_\varphi,b_z$), which reshape the radial potential or add spin-dependent offsets, $a_0$ and $b_0$ enter the spectrum, respectively, as a global energy shift and a spin-momentum coupling. The logic mirrors Sec.~\ref{sec:bounds_spatial}: the Lorentz-violating (LV) contribution to any resolvable spectral feature must be smaller than the experimental resolution $\delta E$.

In the Schrödinger--Pauli limit, $a_0$ appears additively in Eq.~\eqref{spe} with energy dimension. It therefore produces a rigid vertical translation of all levels without changing their spacings. By contrast, $b_0$ enters through the operator
\begin{equation}
\frac{\hbar b_0}{\mu}\,
\boldsymbol{\sigma}\cdot(\mathbf p-q\mathbf A),
\end{equation}
which has energy dimension and couples spin to the kinetic momentum. After separation in cylindrical coordinates this gives $m$-dependent corrections whose magnitude is set by typical azimuthal velocities at the confinement scale $r_0$. In SI units, $a_0$ carries energy dimension (J or eV), while $b_0$ has wavenumber dimension (m$^{-1}$), consistent with Table~\ref{tab:units}.

As in Sec.~\ref{sec:bounds_spatial}, intermediate expressions involve the symmetric-gauge combination
\[
\left(\frac{\hbar m}{r_0} - \frac{eB_0 r_0}{2}\right),
\]
which is nothing but the characteristic azimuthal kinetic momentum at radius $r_0$. When we turn this into a bound, we reinterpret that factor as a gauge-invariant kinetic expectation value. This ensures that the final constraints on $a_0$ and $b_0$ are gauge independent.

Because $a_0$ contributes directly as an additive scalar potential in Eq.~\eqref{spe}, its bound follows immediately:
\begin{equation}
|a_0| \;\lesssim\; \delta E .
\label{bound_a0}
\end{equation}
This limit is universal (it does not depend on device parameters) and improves linearly with spectroscopic resolution.

For $b_0$, we estimate the spin-momentum energy correction as
\begin{equation}
\Delta E_{b_0} 
\sim 
\hbar b_0 \,
\frac{1}{\mu}
\left( 
\frac{\hbar m}{r_0} 
- \frac{eB_0\, r_0}{2} 
\right),
\end{equation}
where $r_0$ is a characteristic lateral scale and $m$ labels the angular-momentum sector resolved experimentally. Requiring $|\Delta E_{b_0}|\le \delta E$ yields
\begin{equation}
|b_0| 
\;\lesssim\;
\frac{\mu \, \delta E}{\hbar^2 \left| \dfrac{m}{r_0} - \dfrac{eB_0 r_0}{2\hbar} \right|}.
\label{bound_b0}
\end{equation}
The right-hand side has units of inverse length, as expected for $b_0$. Unless otherwise stated, we evaluate Eq.~\eqref{bound_b0} using the same GaAs-like reference set adopted in Sec.~\ref{sec:bounds_spatial}:
\[
\mu = 0.067\,m_e,\qquad
r_0 = 10~\text{nm},\qquad
B_0 = 1~\text{T},\qquad
m = 1.
\]

The relevant $\delta E$ values are the spectroscopic resolutions / homogeneous linewidths listed in Table~\ref{tab:linewidths}. Inserting those into Eqs.~\eqref{bound_a0} and \eqref{bound_b0} gives the bounds in Table~\ref{tab:bounds_a0_b0}.

\begin{table}[tbhp]
\centering
\caption{Upper bounds for $a_0$ and $b_0$ inferred from the resolutions in Table~\ref{tab:linewidths}. For $b_0$ we used $\mu=0.067\,m_e$, $r_0=10\,\text{nm}$, $B_0=1\,\text{T}$, $m=1$.}
\begin{tabular}{l@{\hspace{0.3cm}}c@{\hspace{0.3cm}}c@{\hspace{0.3cm}}c}
\toprule
Reference 
& $|a_0|$ (J) 
& $|a_0|$ (eV) 
& $|b_0|$ (m$^{-1}$) \\
\midrule
Vonk et al.~\cite{vonk2021}        
& $1.38 \times 10^{-20}$ 
& $8.61 \times 10^{-2}$ 
& $7.56 \times 10^8$ \\
Stingel et al.~\cite{stingel2023}  
& $4.97 \times 10^{-21}$ 
& $3.10 \times 10^{-2}$ 
& $2.73 \times 10^8$ \\
Gustin et al.~\cite{gustin2021}    
& $2.31 \times 10^{-25}$ 
& $1.44 \times 10^{-6}$ 
& $1.27 \times 10^4$ \\
Pedersen et al.~\cite{pedersen2020}
& $2.40 \times 10^{-25}$ 
& $1.50 \times 10^{-6}$ 
& $1.32 \times 10^4$ \\
L\"obl et al.~\cite{lobl2017}      
& $3.20 \times 10^{-25}$ 
& $2.00 \times 10^{-6}$ 
& $1.76 \times 10^4$ \\
\bottomrule
\end{tabular}
\label{tab:bounds_a0_b0}
\end{table}
Taken together, Tables~\ref{tab:bounds} and \ref{tab:bounds_a0_b0} summarize our experimental \emph{upper bounds} on the minimal SME coefficients. In each case, a background larger than the quoted value would have produced a spin-selective offset, a geometry-dependent spectral distortion, or a uniform level shift exceeding the experimentally resolvable energy scale $\delta E$, and would therefore already have been observable. Smaller backgrounds are not excluded; they remain below the present spectroscopic sensitivity.

\paragraph*{Discussion of trends and robustness.}
(i) The $a_0$ limits are set directly by $\delta E$ and therefore tighten monotonically with improved energy resolution; they do \emph{not} depend on device geometry. Values at the $10^{-25}$\,J scale ($\mu$eV $\sim 10^{-6}$ eV) highlight how competitive solid-state spectroscopy already is for scalar SME sectors.  
(ii) The $b_0$ limits scale with $\mu$, $r_0$, and $B_0$ through Eq.~\eqref{bound_b0}: shrinking $r_0$ or increasing $B_0$ increases the azimuthal kinetic scale in the denominator, tightening the bound. Because $b_0$ couples to an $m$-resolved kinetic expectation, spin- and $m$-selective probes (e.g., polarization-resolved magneto-PL or transport in angular-momentum-filtered states) are the most incisive.  
(iii) Varying $\mu$, $r_0$, $B_0$, and $m$ within realistic III-V ranges changes the $b_0$ numbers by factors of order unity. The qualitative hierarchy remains robust: $|a_0|$ is strongly constrained by $\mu$eV-resolution spectroscopy, while $|b_0|$ requires targeted, spin-resolved, geometry-leveraged measurements to compete.

It is important to emphasize that these condensed-matter bounds are not directly comparable to high-energy compilations, where the minimal SME coefficients are typically quoted with energy dimension for all components in a relativistic description. In our nonrelativistic, effective-mass mapping, $a_0$ keeps energy dimension, but $b_0$ naturally carries inverse length. This reflects the low-energy degrees of freedom and the fact that $b_0$ couples spin to \emph{kinetic} momentum rather than entering as a pure energy offset. The phenomenological message is nevertheless clear: scalar LV backgrounds are strongly constrained by high-resolution level metrology, whereas spin-momentum couplings demand $m$-resolved and geometry-leveraged spectroscopy.

\begin{figure}[tbhp]
\centering
\includegraphics[width=0.48\textwidth]{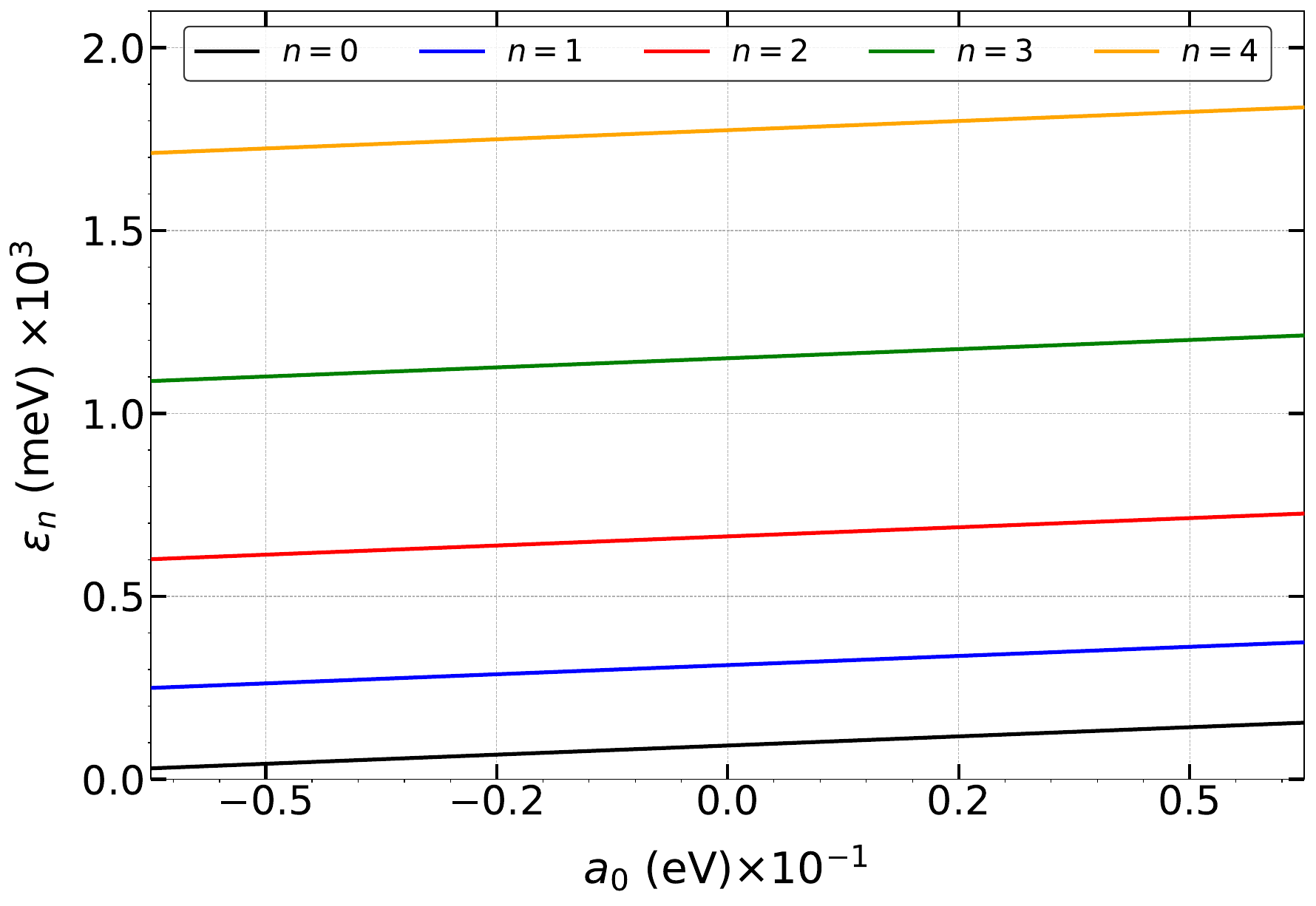}
\caption{Eigenvalues $\varepsilon_n$ (eV) as a function of $a_0$ (eV) for $m=1$, $s=1$. Reference set: $\mu=0.067\,m_e$, $r_0=10$\,nm, $B_0=1$\,T. The range shown corresponds to meV-scale resolutions ($a_0 \sim 10^{-20}$\,eV). The dependence is strictly linear: $a_0$ acts as a uniform vertical shift and does not change level spacings.}
\label{fig:En_vs_a0_case1}
\end{figure}

\begin{figure}[tbhp]
\centering
\includegraphics[width=0.48\textwidth]{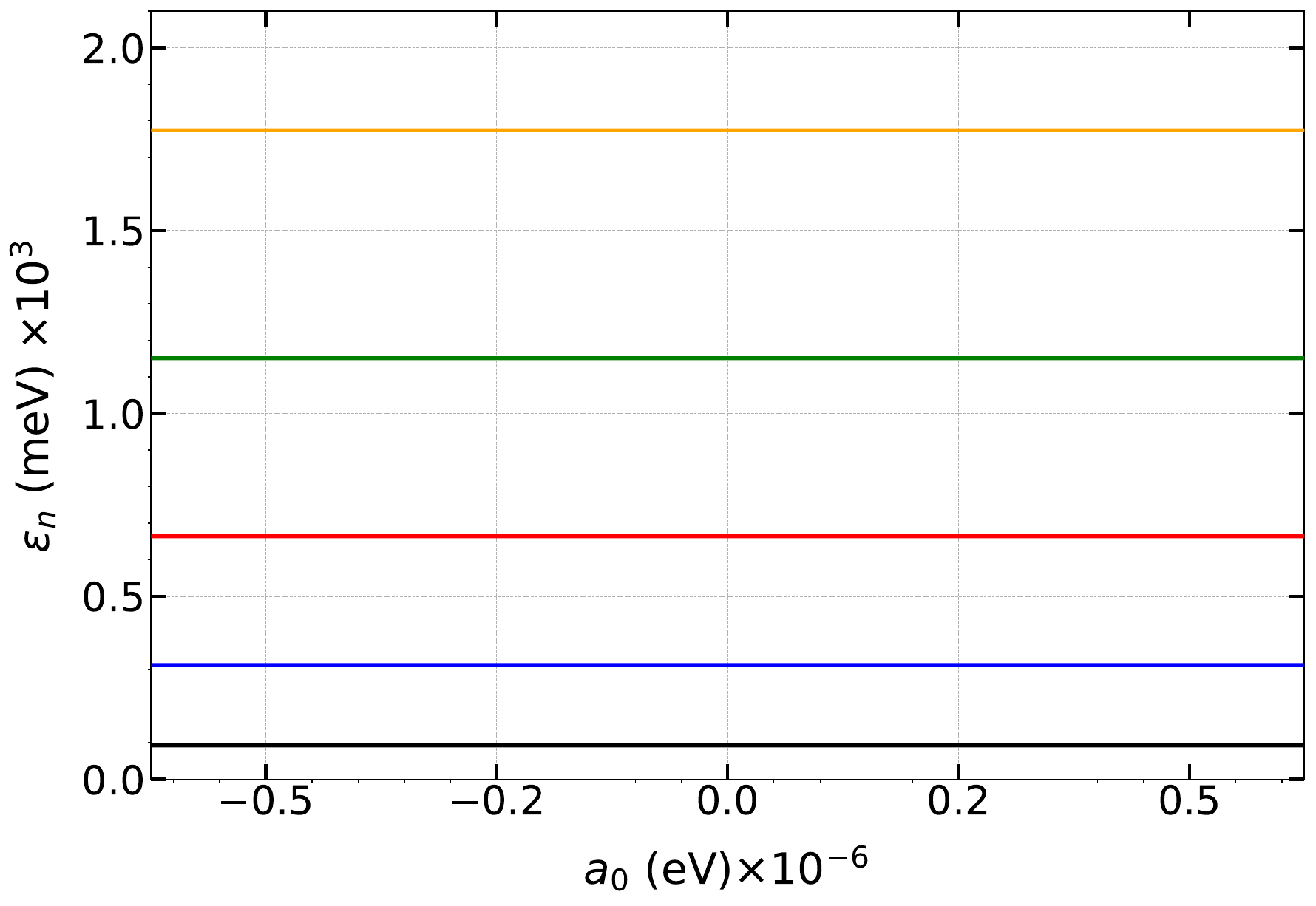}
\caption{Same calculation as Fig.~\ref{fig:En_vs_a0_case1}, now zoomed into the $\mu$eV-resolution regime ($a_0 \sim 10^{-25}$\,eV). The curves remain parallel and offset, confirming that $a_0$ leaves relative splittings essentially unchanged even at the $\mu$eV scale.}
\label{fig:En_vs_a0_case2}
\end{figure}

To isolate the scalar sector for these plots, we set $\mathbf a=\mathbf b=0$ and retained $a_0\neq0$ (and $b_0\neq0$ where indicated). In this configuration, $a_0$ contributes only as a uniform offset to the physical energy $E_n$ without modifying the radial potential. This explains the strictly linear, parallel trends in Figs.~\ref{fig:En_vs_a0_case1}-\ref{fig:En_vs_a0_case2}.

\begin{figure}[tbhp]
\centering
\includegraphics[width=0.48\textwidth]{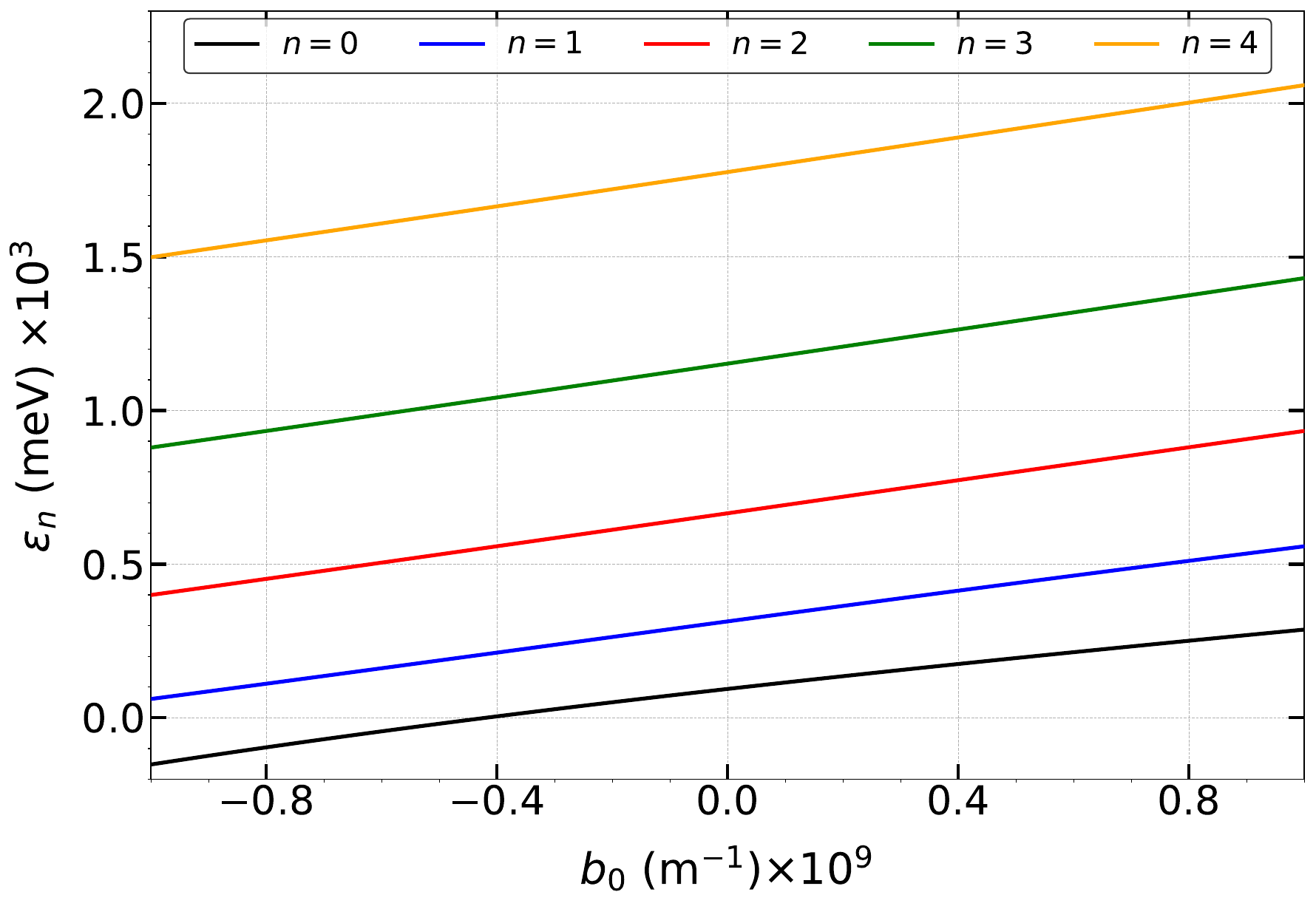}
\caption{Eigenvalues $\varepsilon_n$ plotted versus $b_0$ (m$^{-1}$) for $m=1$, $s=1$, using the parameter range relevant for the bounds at the $10^9$\,m$^{-1}$ level (Table~\ref{tab:bounds_a0_b0}). The vertical axis is rescaled (meV $\times 10^3$) for readability. Within this realistic range, the curves are essentially flat: the $b_0$-induced spin-momentum correction is below current meV-scale resolution. Parameters as in Fig.~\ref{fig:En_vs_a0_case1}.}
\label{fig:Energy_vs_b0_case1}
\end{figure}

\begin{figure}[tbhp]
\centering
\includegraphics[width=0.48\textwidth]{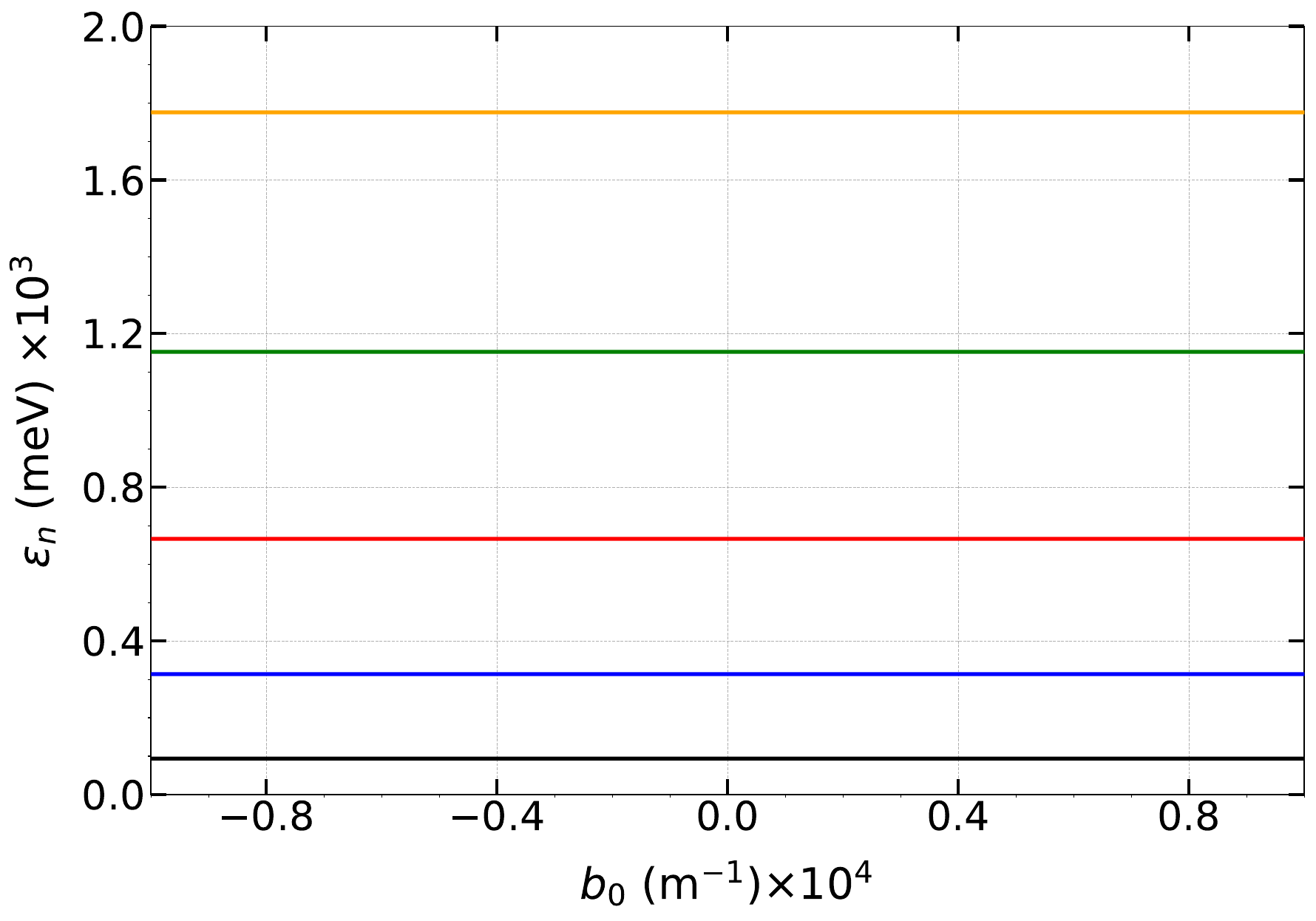}
\caption{Same calculation as Fig.~\ref{fig:Energy_vs_b0_case1}, but zoomed into an order-of-magnitude tighter window ($b_0 \sim 10^4$\,m$^{-1}$), representative of $\mu$eV-level metrology. A weak but visible $n$-dependence now emerges because $b_0$ couples spin to the azimuthal kinetic momentum. Higher-$n$ states, being more spatially extended, are slightly more sensitive.}
\label{fig:Energy_vs_b0_case2}
\end{figure}

For $b_0$, the spin-momentum coupling produces small, $m$-selective distortions of the azimuthal kinetic sector. Within the realistic bounds of Table~\ref{tab:bounds_a0_b0}, these corrections lie below present-day resolution (Fig.~\ref{fig:Energy_vs_b0_case1}). In more optimistic $\mu$eV-scale windows [Fig.~\ref{fig:Energy_vs_b0_case2}], higher-$n$ states--being more extended and thus sampling a slightly larger azimuthal kinetic scale--show an enhanced but still perturbative response. This suggests that experimental strategies targeting excited manifolds, or tuning $(r_0,B_0)$ to increase the azimuthal kinetic factor in Eq.~\eqref{bound_b0}, can optimize sensitivity to $b_0$.

\section{SI-consistent bounds, reporting conventions, and comparison across regimes}
\label{sec:discussion_bounds}

A recurring challenge in the Lorentz-violation (LV) literature is that bounds are quoted across very different kinematic regimes and unit conventions. High-energy and astroparticle searches typically present limits on the minimal SME in terms of parameters with \emph{energy} dimension in a relativistic description. By contrast, our platform is explicitly nonrelativistic (NR): a Schrödinger--Pauli Hamiltonian with effective mass, static electromagnetic fields, and confined two-dimensional electrons. This section consolidates what we have bounded, clarifies unit assignments, and proposes a practical reporting recipe that preserves gauge independence and avoids misleading cross-regime comparisons.

\paragraph*{What is actually bounded here.}
In our NR, effective-mass framework:
\begin{itemize}
\item The scalar coefficients $(a_0,b_0)$ enter, respectively, as a \emph{global energy shift} and a \emph{spin-momentum} coupling in the single-particle Hamiltonian~\eqref{spe}. They are constrained by Eqs.~\eqref{bound_a0} and \eqref{bound_b0} in Sec.~\ref{sec:bounds_scalar}, which translate a measured spectroscopic resolution $\delta E$ into upper limits.
\item Spatial components $(\mathbf a,\mathbf b)$ leave distinct, experimentally identifiable imprints on the radial problem~\eqref{re}:  
  (i) $a_\varphi$ reshapes the effective potential through $1/r$ and $r$ terms and is bounded via the geometry-aware expression in Sec.~\ref{sec:bounds_spatial};  
  (ii) $b_z$ produces a spin-selective offset and is bounded by the universal relation $|b_z|\lesssim\delta E/\hbar$, Eq.~\eqref{eq:bz_bound}.
\end{itemize}
All numerical values in Tables~\ref{tab:bounds} and \ref{tab:bounds_a0_b0} are given in SI units consistent with Table~\ref{tab:units}: $a_0$ in joules (or eV), $a_i$ in kg$\cdot$m/s, $b_0$ in m$^{-1}$, and $b_i$ in s$^{-1}$. These assignments follow directly from the nonrelativistic reduction of the SME and from writing all LV terms at linear order in $a_\mu$ and $b_\mu$. Importantly, the bounds are expressed in terms of \emph{kinetic}, gauge-invariant quantities (e.g., the azimuthal kinetic momentum scale at $r_0$), so the reported limits are not gauge dependent even though we performed the algebra in the symmetric gauge.

\paragraph*{Why direct comparison to relativistic bounds is not straightforward.}
In relativistic SME treatments, all components of $a_\mu$ and $b_\mu$ are often quoted with energy dimension. After performing the nonrelativistic Foldy--Wouthuysen--type expansion and projecting onto a low-energy subspace with effective mass $\mu$, those same background structures reappear in different operator combinations:
\begin{itemize}
\item $a_0$ remains an additive scalar potential (energy dimension).
\item $\mathbf a$ multiplies a velocity operator, so in the NR Hamiltonian it behaves like a momentum-scale background.
\item $\mathbf b$ enters as $\hbar\,\boldsymbol{\sigma}\cdot\mathbf b$, i.e., a spin-dependent frequency shift.
\item $b_0$ multiplies $\boldsymbol{\sigma}\cdot(\mathbf{p}-q\mathbf A)$, which is a spin-momentum coupling; its natural unit in this regime is inverse length.
\end{itemize}
Trying to force all of these back into a single “energy-like” unit requires picking additional, device-specific scales such as a characteristic velocity $v^\star$, radius $r_0$, or cyclotron frequency $\omega_c=eB_0/\mu$. Those scales \emph{do} vary across materials and devices. Consequently, a naive one-line conversion to an ``energy-unit bound'' would silently assume details of a specific platform and risk a misleading comparison to high-energy tables. For this reason, we keep our bounds in SI with their natural NR dimensions, and then describe how to compare across platforms in a controlled way.

\paragraph*{A practical reporting recipe (device-aware, gauge independent).}
For quoting or reusing these bounds on a specific device, we recommend:
\begin{enumerate}
\item Report the SI limit \emph{together with} the spectroscopic resolution used.  
For example: “with $\delta E = 1.5~\mu$eV we obtain $|b_z|\lesssim 2.2\times 10^{9}~\mathrm{s}^{-1}$ via $|b_z|\le\delta E/\hbar$.”
\item For geometry-leveraged coefficients ($a_\varphi$ and $b_0$), explicitly list the device parameters used to evaluate the denominator in Eqs.~\eqref{eq:aphi_bound} and \eqref{bound_b0}, i.e., $(\mu, r_0, B_0, m)$, along with their typical uncertainties or ranges across nominally identical devices.
\item Optionally define \emph{dimensionless} diagnostics that compare LV effects to intrinsic kinetic and instrumental scales without erasing units. Two natural examples are
\begin{align}
\Xi_{a} \equiv \frac{|a_\varphi|}{\mu v_{\varphi}^{\star}},\qquad
\Xi_{b} \equiv \frac{\hbar |b_z|}{\delta E},
\end{align}
where $v_{\varphi}^{\star}$ is the azimuthal velocity scale inferred from the measured spectrum (or from $r_0$ and $\omega_c$). Values $\Xi_{a}\ll 1$ and $\Xi_{b}\ll 1$ indicate that the inferred LV effects are subdominant compared to, respectively, the kinetic scale of the confined electron and the instrumental energy resolution.
\end{enumerate}

\paragraph*{How to place our bounds next to relativistic compilations (with caveats).}
If one nevertheless wants an order-of-magnitude visual comparison to relativistic SME tables, a controlled way is to define \emph{device-tagged energy proxies}:
\begin{align}
\varepsilon^{(a_0)}_{\mathrm{proxy}} &\equiv |a_0|,\\
\varepsilon^{(a_i)}_{\mathrm{proxy}} &\sim |a_i|\, v^{\star},\\
\varepsilon^{(b_i)}_{\mathrm{proxy}} &\sim \hbar\,|b_i|,\\
\varepsilon^{(b_0)}_{\mathrm{proxy}} &\sim \hbar\,|b_0|\, v^{\star},
\end{align}
where $v^{\star}$ is a characteristic group/azimuthal velocity for the probed states (inferred, e.g., from the observed ladder of $m$-resolved levels or from the cyclotron scale $\omega_c$). These $\varepsilon_{\mathrm{proxy}}$ inherit \emph{all} the platform details and must be labeled accordingly (e.g., ``GaAs dot, $r_0=10$~nm, $B_0=1$~T, $\mu=0.067\,m_e$''). They can then be plotted side by side with relativistic bounds \emph{only for qualitative orientation}. Because $v^{\star}$, $r_0$, and $\mu$ change substantially across platforms, such a plot is not a universal ranking; it is a coverage map of which LV operator structures are accessible in which condensed-matter regime.

\paragraph*{Robustness, levers, and outlook.}
Within current $\mu$eV-resolution spectroscopy, the tightest limits arise for $a_0$ (device agnostic, directly set by $\delta E$) and, when geometry is favorable, for $a_\varphi$ [Sec.~\ref{sec:bounds_spatial}]. Axial sectors ($b_z$ and $b_0$) benefit from spin- and $m$-resolved protocols [Secs.~\ref{sec:bounds_spatial} and \ref{sec:bounds_scalar}], and their bounds can be tightened by:  
(i) shrinking $r_0$, which enhances the azimuthal kinetic scale;  
(ii) increasing $B_0$, which boosts $\omega_c$; and  
(iii) targeting excited manifolds, where the effective azimuthal kinetic factor has larger dynamic range.  
Because the formulas are linear in $a_\mu$ and $b_\mu$ and are written in terms of gauge-invariant kinetic quantities, the same procedure carries over directly to other materials (different effective masses, $g$-factors, and $g$-tensors) and to platforms beyond conventional III-V quantum dots or rings.

\paragraph*{Summary for comparison purposes.}
Our SI-consistent, nonrelativistic bounds should be viewed as \emph{device-calibrated constraints} that complement relativistic tables. We provide (i) explicit unit assignments tied to the NR Hamiltonian, (ii) controlled dependence on geometry and field, and (iii) dimensionless diagnostics ($\Xi_a$, $\Xi_b$) that allow cross-platform comparison without hiding assumptions. This preserves physical interpretability and avoids misleading unit conversions, while still allowing qualitative side-by-side plots with high-energy/astroparticle limits when those plots are explicitly annotated with the relevant device parameters and velocity scales.

The bounds reported here have direct physical meaning in the nonrelativistic effective-mass regime. Each coefficient ($a_0$, $a_\varphi$, $b_z$, $b_0$) appears in the Schrödinger--Pauli Hamiltonian as a distinct, experimentally resolvable operator: a uniform scalar shift ($a_0$), a spin-selective offset ($b_z$), a geometry-dependent deformation of the radial confinement ($a_\varphi$), or a spin-momentum coupling ($b_0$). Our limits therefore constrain the maximal strength of each of these operator structures that is still compatible with existing $\mu$eV-resolution spectra. Because the nonrelativistic reduction assigns different physical dimensions to these coefficients (momentum, frequency, inverse length, etc.), the resulting SI bounds are not intended to be compared numerically to relativistic high-energy limits; instead, they characterize how large Lorentz-violating backgrounds could be in a condensed-matter platform without already having been observed.

Beyond their formal connection to the SME, these bounds are of practical interest for condensed-matter platforms themselves: they show that state-of-the-art quantum-dot and ring-like 2D confinement devices, operated with $\mu$eV-resolution magneto-spectroscopy, already act as precision sensors for symmetry-breaking operator structures (uniform scalar shifts, spin-selective offsets, and spin–momentum couplings) that are usually discussed only in high-energy contexts. In this sense, the present analysis frames magnetically confined semiconductor nanostructures not merely as mesoscopic systems, but as tabletop probes of fundamental symmetry.

We emphasize that the present bounds need not be interpreted as direct competitors to ultra-relativistic or astrophysical constraints. There is by now a well-established tradition of using high-precision, low-energy platforms — atomic clocks, trapped ions, comagnetometers — to constrain SME coefficients in situ, without invoking collider-scale energies. Our results place magnetically confined semiconductor nanostructures in this same category: they act as tabletop precision probes of specific Lorentz- and CPT-violating operator structures, with SI-reported sensitivities set by experimentally demonstrated $\mu$eV-scale linewidths. In this sense, the relevance of the bounds is intrinsic to the condensed-matter platform, regardless of whether one extrapolates them to high-energy scenarios.

\section{Projected sensitivities and experimental roadmap}
\label{sec:sensitivity}

The scaling relations derived in Secs.~\ref{sec:bounds_spatial} and \ref{sec:bounds_scalar} make it possible to forecast how the attainable bounds on $(a_0,b_0,a_\varphi,b_z)$ improve with instrumental resolution and device geometry. In the nonrelativistic, effective-mass regime considered here, the parametric behavior is simple:
\begin{align}
|a_0| &\propto \delta E,\\
|b_z| &\propto \delta E/\hbar,\\[4pt]
|a_\varphi| &\propto 
\frac{\mu\,\delta E}{
\big|\,\hbar m/r_0 - (eB_0 r_0)/2\,\big|},\\[4pt]
|b_0| &\propto 
\frac{\mu\,\delta E}{
\hbar^2\big|\,m/r_0 - eB_0 r_0/(2\hbar)\,\big|}.
\end{align}
Here $\delta E$ is the experimentally resolvable energy shift (narrowest homogeneous linewidth); $\mu$ is the effective mass; $r_0$ is a characteristic lateral confinement radius; $B_0$ is the applied perpendicular magnetic field; and $m$ is the azimuthal quantum number of the probed state. As discussed previously, the combinations $\hbar m/r_0 - eB_0 r_0/2$ and $m/r_0 - eB_0 r_0/(2\hbar)$ are shorthand for the azimuthal \emph{kinetic} scale at $r_0$, so the associated bounds are gauge independent.

Two broad trends follow immediately. First, $a_0$ and $b_z$ are ``resolution-limited'': their bounds improve directly with better $\delta E$, with no geometric lever arm. Second, $a_\varphi$ and $b_0$ enjoy geometric leverage: shrinking $r_0$ and increasing $B_0$ both increase the azimuthal kinetic scale, tightening the corresponding bounds for a fixed $\delta E$. Spectroscopy in higher-$|m|$ manifolds offers a similar lever.

Table~\ref{tab:projections} illustrates this leverage for two representative operating points: (i) a conservative ``1~$\mu$eV, $10$~nm, 1~T'' baseline that matches modern quantum-dot spectroscopy, and (ii) a more ambitious ``0.3~$\mu$eV, 8~nm, 3~T'' regime that is within routine reach of resonant, near-transform-limited optical protocols. The $\mu$eV regime has been repeatedly demonstrated in single-dot experiments using resonant (or quasi-resonant) excitation~\cite{gustin2021,pedersen2020,lobl2017,laferriere2023}, so these projections correspond to realistic near-term performance.

\begin{table*}[tbhp]
\centering
\caption{Projected 95\% CL-style sensitivities for the SME coefficients, obtained by varying spectroscopic resolution $\delta E$ and device geometry. We assume $\mu=0.067\,m_e$ for concreteness. The $\mu$eV-scale resolutions and field/size combinations are representative of near-transform-limited single-dot measurements reported in Refs.~\cite{gustin2021,pedersen2020,lobl2017,laferriere2023}. ``$\sim 3\times$ tighter'' etc.\ indicate the relative improvement over the baseline geometry due to enhanced azimuthal kinetic scale.}
\label{tab:projections}
\begin{tabular}{lcccc}
\toprule
$(\delta E, r_0, B_0)$ 
& $|a_0|$ (eV) 
& $|b_z|$ (s$^{-1}$) 
& $|a_\varphi|$ (kg\,m/s) 
& $|b_0|$ (m$^{-1}$)\\
\midrule
$1~\mu$eV,\; $10$\,nm,\; $1$\,T 
& $1.0\times10^{-6}$ 
& $1.5\times10^{9}$ 
& $1.5\times10^{-30}$ 
& $1.3\times10^{4}$\\[3pt]
$0.3~\mu$eV,\; $8$\,nm,\; $3$\,T 
& $3.0\times10^{-7}$ 
& $4.5\times10^{8}$ 
& $\sim 3\times$ tighter 
& $\sim 4\times$ tighter\\
\bottomrule
\end{tabular}
\end{table*}

In parallel with raw sensitivity, it is essential to control systematic shifts and drifts at or below the target $\mu$eV scale. Table~\ref{tab:errors} summarizes a representative uncertainty budget for polarization- and $m$-resolved spectroscopy in quantum-dot platforms, along with standard mitigation strategies. These procedures, laser stabilization, careful $B_0$ calibration, Stark-shift compensation, suppression of spectral diffusion, and model-selection diagnostics for line fits, are already routine in near-transform-limited experiments~\cite{gustin2021,pedersen2020,laferriere2023}.

\begin{table*}[tbhp]
\centering
\caption{Representative uncertainty budget for polarization- and $m$-resolved spectroscopy, together with standard mitigation strategies in near-transform-limited quantum-dot experiments~\cite{gustin2021,pedersen2020,laferriere2023}. ``AIC/BIC'' refers to standard Akaike/Bayesian information-criterion model selection between, e.g., Voigt and Lorentzian fits.}
\label{tab:errors}
\begin{tabular}{lcc}
\toprule
Source & Impact on $E$ & Mitigation \\
\midrule
Laser drift 
& $\lesssim 0.3~\mu$eV 
& frequency-locked laser / Fabry-P\'erot reference \\
$B_0$ calibration 
& slope bias in $E(B_0)$ 
& Hall/ESR cross-calibration \\
Stark drift 
& centroid shift 
& gate mapping, extrapolation to $V\to0$ \\
Spectral diffusion 
& inhomogeneous broadening 
& diode p-i-n structures, slow scans, resonant driving \\
Fit model choice 
& centroid bias 
& Voigt vs.\ Lorentz; AIC/BIC model selection \\
\bottomrule
\end{tabular}
\end{table*}

Operationally, these scalings suggest a concrete experimental roadmap:

\begin{itemize}
\item \textbf{Axial sector ($b_z$ and $b_0$).}  
Spin- and $m$-resolved magneto-photoluminescence (or resonance fluorescence) directly targets the axial LV backgrounds. The polarization-resolved extrapolation of $E(B_0)$ as $B_0\to0$ isolates $b_z$ via the spin-selective offset $s\hbar b_z$. A systematic sweep over angular-momentum sectors and device sizes then parametrizes the regressor
\[
X \equiv \left(\frac{m}{r_0} - \frac{eB_0 r_0}{2\hbar}\right),
\]
which is the kinetic prefactor entering the $b_0$ bound in Eq.~\eqref{bound_b0}. Arrays of nominally identical dots or rings with $r_0\in[8,30]$\,nm, combined with fields up to a few tesla, provide enough geometric leverage to disentangle $b_0$ from pure resolution limits.

Operating near the condition 
\[
\frac{eB_0 r_0^2}{2\hbar} \approx m
\]
maximizes the dynamic range of $X$, and thus the statistical power of a global fit for $b_0$.

\item \textbf{Scalar sector ($a_0$).}  
Because $a_0$ produces a rigid vertical energy shift [Eq.~\eqref{bound_a0}], it is best constrained simply by pushing $\delta E$ down. State-of-the-art single-dot linewidths in the $\mu$eV range already enforce $|a_0|$ at the $10^{-25}$~J ($10^{-6}$~eV) level, see Table~\ref{tab:bounds_a0_b0}.

\item \textbf{Azimuthal spatial sector ($a_\varphi$).}  
Bounds on $a_\varphi$ improve with smaller $r_0$, higher $B_0$, and larger $|m|$, cf.\ Eq.~\eqref{eq:aphi_bound}. In practice, this means that lithographic control of lateral size and the ability to probe well-defined angular-momentum manifolds are as important as raw spectral resolution. Designs that confine carriers in narrow rings or small-radius dots at multi-tesla fields are particularly advantageous.
\end{itemize}

In summary, the same spectroscopy tools that are already standard for single quantum dots and mesoscopic rings (polarization-resolved PL or resonance fluorescence under tunable $B_0$) are sufficient, in principle, to access all four minimal SME coefficients considered here. The roadmap above turns the scaling relations of Secs.~\ref{sec:bounds_spatial} and \ref{sec:bounds_scalar} into an experimental program: improve $\delta E$ for $a_0$ and $b_z$, engineer $(r_0,B_0,m)$ leverage for $a_\varphi$ and $b_0$, and control the $\mu$eV-level systematics summarized in Table~\ref{tab:errors}.

\section{Experimental systematics, gauge independence, and cross-platform diagnostics}
\label{sec:systematics}

\subsection{Systematics and uncertainty budget}

Because the bounds we report are linear functions of measured level shifts (Secs.~\ref{sec:bounds_spatial} and \ref{sec:bounds_scalar}), controlling conventional sources of drift and broadening is essential. Polarization-swap sequences with randomized ordering, $B_0$ reversals, and gate-voltage sweeps provide robust null tests to distinguish genuine spin-selective offsets from ordinary Zeeman and Stark responses. Cross-geometry consistency across devices with different $r_0$ tests whether any putative LV signal respects the geometric scaling predicted in Sec.~\ref{sec:sensitivity}.

In practice, near-transform-limited excitation protocols~\cite{pedersen2020,laferriere2023} and stabilized resonant (or quasi-resonant) addressing~\cite{gustin2021} suppress power broadening and spectral wandering down to the $\mu$eV range. Independent calibration of $B_0$ (via a Hall probe or ESR reference) removes slope biases in $E(B_0)$ that would otherwise masquerade as $b_z$. The uncertainty budget summarized in Table~\ref{tab:errors} (see Sec.~\ref{sec:sensitivity}) lists the dominant technical systematics, their typical impact on extracted line centers, and common mitigation strategies (laser frequency locking to a Fabry-P\'erot reference, Stark-shift extrapolation to $V\to0$, suppression of charge noise and spectral diffusion in diode p-i-n structures, and AIC/BIC-guided line-shape model selection). Our bounds propagate statistical and systematic components in quadrature.

\subsection{Gauge independence: an operational statement}
\label{sec:gauge}

Although the derivations in Sec.~\ref{sec:hamiltonian} employed the symmetric gauge for notational convenience, the final bounds are expressed in terms of \emph{gauge-invariant} kinetic quantities. Under a gauge transformation $\mathbf{A}\rightarrow\mathbf{A}+\nabla\chi$, the canonical momentum changes, but the kinetic momentum
\[
\boldsymbol{\pi} = \mathbf{p}-q\mathbf{A}
\]
and the associated velocity $\mathbf{v} \propto \boldsymbol{\pi}/\mu$ remain invariant and control the physical spectrum.

The geometry-aware denominators appearing in the bounds for $a_\varphi$ and $b_0$ [Eqs.~\eqref{eq:aphi_bound} and \eqref{bound_b0}] can be written as expectation values of the azimuthal component of the kinetic momentum, $\pi_\varphi$, divided by $\mu$ and evaluated at the characteristic confinement radius $r_0$. In shorthand, this shows up in the symmetric-gauge algebra as
\[
\left(
\frac{\hbar m}{r_0}
-
\frac{eB_0 r_0}{2}
\right)
\quad\text{or}\quad
\left(
\frac{m}{r_0}
-
\frac{eB_0 r_0}{2\hbar}
\right),
\]
but physically these are just the azimuthal kinetic scales that enter the measurable spectrum. Recasting the bounds in terms of those kinetic scales makes them manifestly gauge independent. This is why the limits quoted in Tables~\ref{tab:bounds} and \ref{tab:bounds_a0_b0} are not artifacts of any particular vector potential.

\subsection{Dimensionless diagnostics for cross-platform comparison}
\label{sec:dimensionless}

To enable side-by-side comparison across different materials and device geometries--without forcing ad-hoc energy conversions--we complement the SI bounds with dimensionless diagnostics that normalize LV effects to natural kinetic or instrumental scales:
\begin{equation}
\Xi_a
=
\frac{|a_\varphi|}{\mu\, v_\varphi^\star},
\qquad
\Xi_b
=
\frac{\hbar |b_z|}{\delta E}.
\end{equation}
Here $v_\varphi^\star$ denotes an azimuthal velocity scale inferred from the measured spectrum (for example, from the $m$-resolved ladder spacings) or from a semiclassical estimate
\[
v_\varphi^\star
\simeq
\left|
\frac{\hbar m}{\mu r_0}
-
\frac{eB_0 r_0}{2\mu}
\right|.
\]
By construction, $\Xi_a \ll 1$ signals that spatial-vector LV effects (here encoded in $a_\varphi$) are subdominant to the kinetic scale of the confined electron, whereas $\Xi_b \ll 1$ guarantees that axial LV shifts (here set by $b_z$) remain below the instrumental resolution floor $\delta E$.

For experiments targeting $b_z$, polarization-resolved fits of $E(B_0)$ in the $\mu$eV regime typically yield $\Xi_b \lesssim 1$ even when $|b_z|$ is quoted only as an upper limit. This directly shows how close the experiment operates to the detection floor. For geometry-leveraged coefficients such as $a_\varphi$, $\Xi_a$ makes explicit the available levers $(r_0,B_0,m,\mu)$: shrinking $r_0$ or increasing $B_0$ raises $v_\varphi^\star$, which tightens the \emph{same} raw SI upper bound into a smaller $\Xi_a$.

One may extend this logic to $b_0$ and $a_0$ by defining
\[
\tilde\Xi_{b_0}
\equiv
\frac{\hbar\,|b_0|\,v_\varphi^\star}{\delta E},
\qquad
\tilde\Xi_{a_0}
\equiv
\frac{|a_0|}{\delta E}.
\]
These quantify how close the spin-momentum coupling $b_0$ and the uniform shift $a_0$ are to the experimental floor. We emphasize that $\Xi_a$, $\Xi_b$, $\tilde\Xi_{b_0}$, and $\tilde\Xi_{a_0}$ are not substitutes for SI reporting; rather, they act as compact summary metrics for planning upgrades (smaller $r_0$, higher $B_0$, narrower $\delta E$) and for comparing different material platforms on equal footing.

\subsection{Data, code, and numerical robustness}
\label{sec:data_code_robustness}

All analysis required to extract $b_z$ and $b_0$ from polarization- and $m$-resolved spectroscopy is provided as Supplemental Material. The package includes:  
(i) a self-contained Python script that performs weighted linear regressions with full uncertainty propagation;  
(ii) a CSV template specifying the required columns (line centers and $1\sigma$ errors in eV, device parameters, and $B_0$); and  
(iii) example datasets reproducing the trends discussed in Sec.~\ref{sec:sensitivity}.

Operationally, the script converts measured line centers (and their uncertainties) into SI units, performs independent fits for $\sigma^\pm$ polarizations to extract the $B_0\to0$ intercepts (hence $b_z$), and regresses the transition energy $E$ versus 
\[
X \equiv \frac{m}{r_0} - \frac{eB_0 r_0}{2\hbar}
\]
to obtain $b_0 = \mu C/\hbar^2$, where $C$ is the fitted slope. The code returns best-fit values and 95\% confidence intervals, produces diagnostic plots, and writes a machine-readable JSON summary for reproducibility.

On the modeling side, the finite-difference solution of the radial eigenproblem~\eqref{re} is numerically stable across the full parameter range we consider. We varied the radial grid size from $N = 1\times10^{4}$ to $4\times10^{4}$, the inner cutoff from $r_{\min}=10^{-7}r_0$ to $10^{-5}r_0$, and the outer boundary from $r_{\max}=10r_0$ to $15r_0$, observing changes below $10^{-6}$\,meV for the ten lowest eigenvalues. The inferred bounds shifted by less than $0.5\%$, well below experimental uncertainties. Cross-checks using alternative boundary conditions (e.g., Neumann at $r_{\min}$) and a standard shooting method produced spectra within the same tolerance. This confirms that our conclusions are limited by experimental resolution and systematics, not by numerical artifacts.

For convenience and to aid adoption by other groups, we also supply a minimal ``sensitivity worksheet'' in the Supplemental Material. Given target values $(\delta E,r_0,B_0,\mu,m)$, the worksheet evaluates the projected bounds appearing in Table~\ref{tab:projections}. The entries are calibrated to $\mu$eV-scale, near-transform-limited quantum-dot spectroscopy~\cite{gustin2021,pedersen2020,lobl2017,laferriere2023}, so that prospective experiments can read off realistic performance targets directly against demonstrated capabilities.

\section{Conclusions}
\label{concl}

We have presented an SI-consistent, gauge-independent framework to test Lorentz- and CPT-violating backgrounds in magnetically confined two-dimensional electron systems within the minimal SME, working in the nonrelativistic (Schrödinger--Pauli) limit with effective mass. By deriving the cylindrically symmetric radial eigenproblem [Eq.~\eqref{re}] from the modified single-particle Hamiltonian [Eq.~\eqref{spe}], we made explicit how each SME coefficient maps onto an experimentally accessible spectral signature:

\begin{itemize}
\item The scalar coefficients $a_0$ and $b_0$ act, respectively, as a uniform energy shift and as a spin-momentum coupling. Their bounds [Eqs.~\eqref{bound_a0} and \eqref{bound_b0}, Table~\ref{tab:bounds_a0_b0}] follow directly from the spectroscopic resolution $\delta E$ and from the azimuthal kinetic scale at the device radius $r_0$.

\item The spatial projections $a_\varphi$ and $b_z$ reshape, respectively, the effective radial potential (through $1/r$ and $r$ terms) and the spin-resolved energy offset. Their bounds [Eqs.~\eqref{eq:aphi_bound} and \eqref{eq:bz_bound}, Table~\ref{tab:bounds}] are controlled either by geometry $(r_0,B_0,m,\mu)$ or purely by $\delta E$.

\end{itemize}

From these structures we obtained compact phenomenological bounds that translate state-of-the-art linewidths and device parameters directly into limits on minimal SME coefficients, expressed in SI units. Universal relations set limits for $a_0$ and $b_z$ purely from the resolution scale, whereas the bounds for $a_\varphi$ and $b_0$ expose explicit \emph{levers}--lateral size $r_0$, field strength $B_0$, and angular-momentum selectivity--that can be engineered to tighten sensitivity. Because the relevant combinations are ultimately gauge-invariant kinetic quantities, the final limits are not tied to any particular gauge choice.

Finite-difference numerics of the radial equation corroborate the analytical scaling laws and clarify the spectral fingerprints:  
(1) $a_0$ produces parallel, uniformly shifted level manifolds;  
(2) $b_z$ appears as a spin-selective offset proportional to $s\hbar b_z$;  
(3) $a_\varphi$ deforms the effective confinement via Coulomb-like ($1/r$) and linear-in-$r$ terms; and  
(4) $b_0$ yields weak, $m$-dependent spin-momentum corrections that become more visible in spin- and polarization-resolved spectroscopy and in higher excited states. Within currently demonstrated $\mu$eV resolutions, the tightest constraints arise for $a_0$ and, when geometry is favorable, for $a_\varphi$, while axial sectors benefit from targeted spin and angular-momentum selectivity.

Conceptually, this work clarifies dimensional assignments and gauge issues for condensed-matter realizations of the SME: spatial $a_i$ carry momentum dimension and axial $b_i$ carry inverse-time/length dimensions in the effective-mass description, rather than being all quoted in ``energy units'' as in relativistic tables. We also introduced dimensionless diagnostics ($\Xi_a$, $\Xi_b$, and their extensions) that summarize sensitivity relative to kinetic and instrumental scales, enabling meaningful cross-platform comparison without erasing material- and geometry-specific information.

Natural next steps include: relaxing idealizations (many-body effects, hyperfine and spin-orbit interactions, disorder, phonons); exploring materials with distinct effective masses and $g$-factors; exploiting level crossings or anticrossings and time-dependent control to accumulate geometric and dynamical phases; and extending the same pipeline to nonminimal SME operators. Altogether, magnetically confined 2D semiconductor platforms emerge as precise and versatile laboratories in which to confront Lorentz symmetry at low energies, complementing high-energy and astroparticle searches with device-level metrology.

\section*{Acknowledgments}

This work was supported by CAPES (Finance Code 001), CNPq/PQ/306308/2022-3, and FAPEMA/Universal/06395/22.

\bibliographystyle{apsrev4-2}
%\bibliography{References}

\end{document}